\documentclass{aa}

\usepackage{graphicx}
\usepackage{txfonts}
\usepackage{siunitx}

\newcommand{\kms}{{km\,s}$^{-1}$}

\newcommand{\teff}{$T_\mathrm{eff}$\,}
\newcommand{\logg}{$\log g$\,}
\newcommand{\vt}{$v_{\rm{t}}$}
\newcommand{\vsini}{$v_{\rm{e}} \sin i$}

\begin{document}

   \title{Hubble spectroscopy of LB-1: Comparison with B+black-hole and Be+stripped-star models.\thanks{Table 2 is only available in electronic form
                at the CDS via anonymous ftp to cdsarc.u-strasbg.fr (130.79.128.5)
                or via http://cdsweb.u-strasbg.fr/cgi-bin/qcat?J/A+A/
   }}

   \titlerunning{Hubble spectroscopy of LB-1}


   \author{D.~J. Lennon\inst{1,2},
          J. Ma\'iz Apell\'aniz\inst{3},
          A. Irrgang\inst{4},
          R. Bohlin\inst{5},
          S. Deustua\inst{5},
          P. L. Dufton\inst{6},
          S. Sim\'on-D\'iaz\inst{1,2}
          A. Herrero\inst{1,2},
          J. Casares\inst{1,2},
          T. Mu\~noz-Darias\inst{1,2},
          S. J. Smartt\inst{6},
          J.~I. Gonz\'alez Hern\'andez\inst{1,2},
          A. de Burgos\inst{1,2}
          }
\authorrunning{D.~J.~Lennon et al.} 

   \institute{Instituto de Astrof\'isica de Canarias,E-\num[detect-all]{38200} La Laguna, Tenerife, Spain
         \and
              Dpto. Astrof\'isica, Universidad de La Laguna, E-\num[detect-all]{38205} La Laguna, Tenerife, Spain
         \and
             Centro de Astrobiolog\'ia, ESAC campus, Villanueva de la Ca\~nada, E-\num[detect-all]{28692}, Spain
          \and
             Dr. Karl Remeis-Observatory \& ECAP, Astronomical Institute, Friedrich-Alexander University Erlangen-Nuremberg (FAU), Sternwartstr. 7, \num[detect-all]{96049} Bamberg, Germany
         \and
             Space Telescope Science Institute, 3700 San Martin Drive, Baltimore, MD \num[detect-all]{21218}, USA
        \and
            Astrophysics Research Centre, School of Mathematics \& Physics,  Queen’s University, Belfast, BT7 1NN, UK 
             }

   \date{}

 
  \abstract
  {LB-1 (alias ALS\,8775) has  been proposed as either an X-ray dim B-type star plus black hole (B+BH) binary or a Be star plus an inflated stripped star (Be+Bstr) binary. The latter hypothesis contingent upon the detection and characterization of the hidden broad-lined star in a composite optical spectrum.} 
%
  {Our study is aimed at testing the published B+BH (single star) and Be+Bstr (binary star) models using a flux-calibrated UV-optical-IR spectrum.} 
   {The Space Telescope Imaging Spectrograph (STIS) on board the Hubble Space Telescope (HST) was used to obtain a flux-calibrated spectrum with an accuracy of $\sim$1\%. We compared these data with non-local thermal equilibrium (non-LTE) spectral energy distributions (SED) and line profiles for the proposed models. The Hubble data, together with the $Gaia$ EDR3 parallax and a well-determined extinction, were used to provide tight constraints on the properties and stellar luminosities of the LB-1 system.
   In the case of the Be+Bstr model we adopted the published flux ratio for the Be and Bstr stars, re-determined the \teff\ of the Bstr using the silicon ionization balance, and inferred \teff\ for the Be star from the fit to the SED.}
   {The UV data strongly constrain the microturbulence velocity to $\lesssim$2\,\kms for the stellar components of both models. We also find stellar parameters consistent with previous results, but with greater precision enabled by the Hubble SED.
   For the B+BH single-star model, we find the parameters (\teff, $\log(L/{\rm L}_\odot)$, $M_{\rm spec}$/M$_\odot$) of the B-type star to be 
   (\num{15300}$\pm300$\,K, $3.23^{+0.09}_{-0.10}$, $5.2^{+1.8}_{-1.4}$).
   For the Bstr star we obtain 
   (\num{12500}$\pm100$\,K, $2.70^{+0.09}_{-0.09}$, $0.8^{+0.5}_{-0.3}$), and for the Be star
   (\num{18900}$\pm200$\,K, $3.04^{+0.09}_{-0.09}$, $3.4^{+3.5}_{-1.8}$).
   While the Be+Bstr model is a better fit to the \ion{He}{I} lines and cores of the Balmer lines in the optical, the B+BH model provides a better fit to the \ion{Si}{IV} resonance lines in the UV.  The analysis also implies that
   the Bstr star has roughly twice the solar silicon abundance, which is difficult to reconcile with a stripped star origin.
   The Be star, on the other hand, has a rather low luminosity 
   and a spectroscopic mass that is inconsistent with its possible dynamical mass. 
  }
   {We provide tight constraints on the stellar luminosities of the Be+Bstr and B+BH models. For the former, the Bstr star appears to be silicon-rich, while the notional Be star appears to be sub-luminous for a classical Be star of its temperature and the predicted UV spectrum is inconsistent with the data. This latter issue can be significantly improved by reducing the \teff\ and radius of the Be star, at the cost, however, of a different mass ratio as a result. In the B+BH model, the single B-type spectrum is a good match to the UV spectrum. Adopting a mass ratio of $5.1\pm0.1,$ from the literature, implies a BH mass of $\sim21^{+9}_{-8}\,M_\odot$. }

   \keywords{techniques: spectroscopic, binaries: spectroscopic, stars: black holes, stars: early-type, stars: fundamental parameters}

   \maketitle
%

\section{Introduction}

\citet{liu} reported LB-1 (alias ALS~8775) as a long-period B-type star plus black hole (B+BH) binary composed of an $\sim$8 M$_\odot$ star and a $\sim$70\,M$_\odot$ BH. This model was subsequently revised and added to by a number of authors and the current state of play is nicely summarized by \citet{liu2020}.  Broadly speaking, models depend on whether or not the small apparent motions of the strong Balmer emission lines originating from a disk associated with either a Be star or a BH are taken to represent the reflex motion of that companion around the narrow-lined B-type star.  
This paper examines two competing models. First, there is the B+BH model, where the mass of the B-type star is 3--5\,M$_\odot$ \citep{sergio2020,abdulmasih}, with a corresponding reduction in the potential BH mass and the Be star plus stripped helium star model  \cite[Be+Bstr:][]{shenar}. 
In the latter model, the system is composed of two stellar components of approximately equal visual magnitudes, the Be star being hidden in the optical spectrum due to its high projected rotational velocity (\vsini). The large radius of the B-type stripped star is a consequence of its blue-ward evolution in the Hertzsprung-Russell diagram (HRD) on its way to the He burning main sequence following mass transfer to the Be-type star. This also explains the low \vsini\ of the stripped star, as the previous mass donor, and the high \vsini\ of the Be star, as the mass recipient. 
A similar model \citep{bodensteiner2020} has been put forward for the system HR\,6819, which \citet{rivinius} argued contained a B+BH binary with a tertiary (Be) star in a very long-period orbit. 

This paper presents new flux-calibrated low and medium resolution Hubble spectroscopy covering far UV to near-IR wavelengths that, combined with the recent $Gaia$ EDR3 parallax, provide new constraints on the properties of the proposed models.

   \begin{figure*}
   \centering
   \includegraphics[width=0.965\hsize]{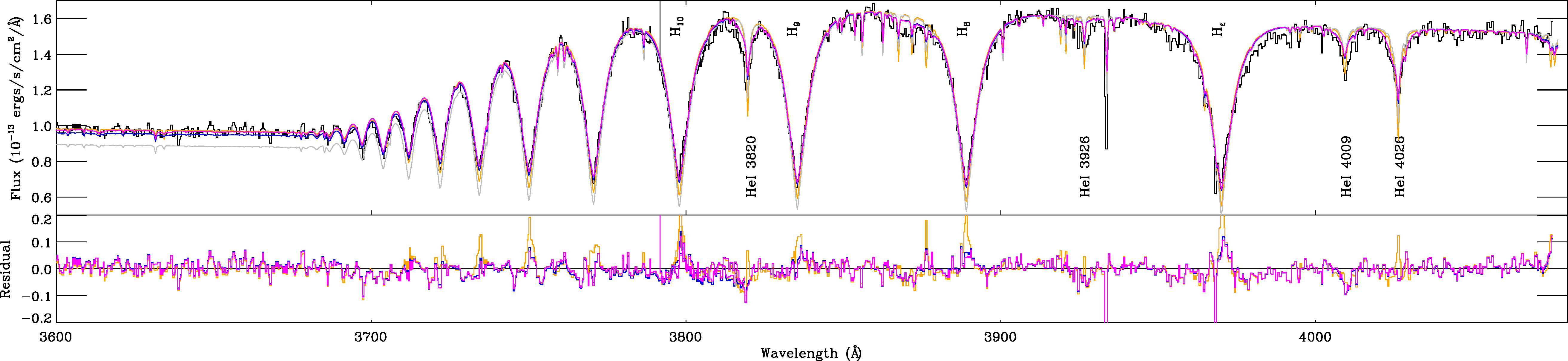}
   \includegraphics[width=1.0\hsize]{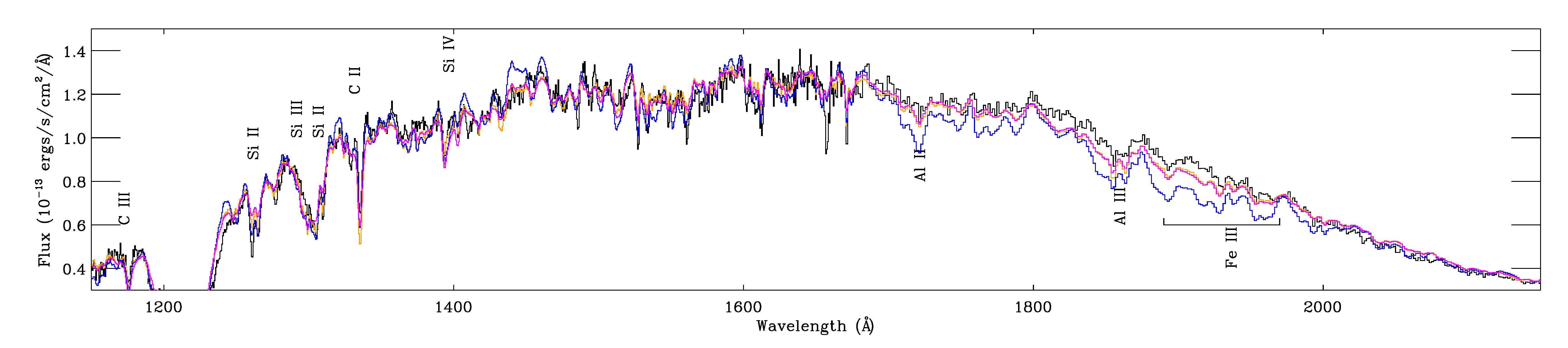}
      \caption{
      Comparison of Hubble data (black) with single B-type star model (orange) and binary Be+Bstr model (magenta), as summarized in Table~\ref{tab:results}, with the main contributing species labeling specific absorption features. The top panel illustrates the good fit of both models to the G430M data, with the binary model being a better fit to the Hydrogen line cores and to the \ion{He}{I} 4026\,\AA\ line, as indicated by the residuals plotted below the spectra. Also shown are the models of \citet[][grey]{sergio2020} and \citet[][blue]{shenar}. The lower panel compares the FUV data (same color coding) illustrating that, for example, the \ion{Fe}{III} complex of lines around 1900\,\AA\ is much too strong in the \citet{shenar} model. Neither interstellar lines nor disk emission are included in the models.
              }
         \label{fig:compare}
   \end{figure*}

\begin{table}[ht]
\caption{Summary of Hubble spectroscopic data obtained with STIS and WFC3/IR under program 16079, indicating the relevant dataset unique identifier, spectral element, aperture, exposure time, and wavelength coverage.}
\centering
\begin{tabular}{lllrl}
\hline
\hline
Dataset & Spectral &  Aper. & Exp.& Wavelength \\
Ident. & Element & & (s) & range (\AA) \\ \hline
oe9l02010 & G430M & 52x0.1 & 242 & 3537--3823 \\
oe9l02020 & G430M & 52x0.1 & 142 & 3793--4079 \\ 
oe9l02030 & G430L & 52x2 & 264 & 2900--5700 \\
oe9l02040 & G750L & 52x2 & 204 & 5240--\num{10270} \\
oe9l01010 & G140L & 52x2 & 1040 & 1150--1730 \\
oe9l01020 & G230L & 52x2 & 840 & 1570--3180 \\
ie9l03010 & G102 & N/A & 33.3 & 8000--\num{11500} \\
ie9l03020 & G141 & N/A & 33.3 & \num{10750}--\num{17000} \\
\hline
\end{tabular}
\label{table:data}
\end{table}

\section{Observational data}

The spectroscopic data were obtained under program GO\,16079 (P.I. Lennon) in response to the Cycle 27 mid-cycle call for proposals.
The primary objective was to obtain a flux calibrated spectral energy distribution (SED) with an accuracy of 1\%\ or better \citep{bohlin2020}, following the logic set out in \citet{bohlin2014}. 
We therefore obtained low-resolution wide slit spectrophotometry in the UV and optical using the Space Telescope Imaging Spectrograph (STIS) and near-IR grism spectra obtained with the Wide Field Camera 3 (WFC3). In addition, we obtained narrow-slit medium resolution spectra covering the Balmer jump and the highest members of the Balmer series of lines from H$_\epsilon$ to the series limit. 
Data were processed using Instrument Development Team (IDL) programs that enable a superior flux calibration compared to the standard pipeline products \citep{fitzpatrick2019} and is identical to the methods used to process Hubble flux standards \citep{bohlin2019,bohlinIR}.

A summary of the spectroscopic observations is provided in Table \ref{table:data}, while the merged SED is attached as an ASCII table, Table \ref{table:sed}, while the full version is available electronically. 
As expected for the low-resolution data, small offsets of the wavelength scale of $\sim$1 pixel are apparent, due to slight off-centering in the wide slit. These corrections were determined by comparison with the predicted radial velocity of the narrow-lined star using the ephemeris of \citet{liu2020} and have been applied to the SED.
The G430M data were taken with a narrow slit and before use were corrected for slit losses by applying scalar correction factors of 1.123 and 1.150 to the 3537--3823 and 3793--4079\,\AA\ data, respectively, as determined from the low-resolution data. 

We also obtained short WFC3/IR images, using the on/off-band P$\beta$ filters (F128N/F130N) to check for possible image extension due to circumstellar emission or potential blending of a close optical companion.  The PSFs of LB-1 appear point-like in both images, with full-width half-maxima similar to each other and to a nearby faint field star with FWHM=0.23" and ellipticities of 0.3\% and 1.3\% for the F128N and F130N filters, respectively.

\begin{table*}[h]
    \centering
    \caption{Flux-calibrated spectrum. The complete version of this table, together with a description of the columns, is available online as an ASCII table.} 
    \begin{tabular}{llllllll} \hline \hline
     Wavelength & Flux & Stat-Error & Sys-error & FWHM & Quality & Time\\
     \hline
 \num{1140.400146} & 3.0854E-14 & 1.0286E-14 & 3.0854E-16  &   1.167 &  1 &   1040.0 \\
 \num{1140.983887} & 3.0082E-14 & 9.2224E-15 & 3.0082E-16  &   1.167 &  1 &   1040.0 \\
 \num{1141.567627} & 3.2377E-14 & 8.8605E-15 & 3.2377E-16  &   1.167 &  1 &   1040.0 \\
 ... & & & & & &   \\
 \hline
   \end{tabular}
    \label{table:sed}
\end{table*}

\section{Methods}

The data were modeled using the solar metallicity grid of TLUSTY plane parallel non-LTE model atmospheres for B-type stars \citep{lanz2007}, with a microturbulent velocity that is appropriate for main-sequence stars (\vt=2\,\kms). These were supplemented with additional models as required, using the TLUSTY model atmosphere code \citep{hubeny1988,hubeny1995} and with the same input model atoms as \citet{lanz2007}. 
We tested two scenarios, the B+BH scenario as represented by a "single" stellar model \citep{sergio2020}, and the Be+Bstr scenario represented by a "binary" composite of two models \citep{shenar}. Synthetic spectra were convolved with appropriate line spread functions (LSF) taken from the STIS web pages\footnote{https://www.stsci.edu/hst/instrumentation/stis/performance/spectral-resolution} prior to re-binning them to match the data. We adopted vacuum wavelengths to match the pipeline data products  and radial velocity shifts appropriate to the phase of each observation were taken from the ephemeris of \citet{liu2020}. 

The primary diagnostics feature the Balmer jump (or Balmer decrement: BD), Balmer line profiles (in the G430M data), and strong metal lines in the UV (Figure 1, lower panel). The SED constrains the extinction law that is required to fit the flux calibrated data, which, together with the $Gaia$ EDR3 parallax measurement, enable a precise estimate of stellar angular radii, the ratio of the stellar radius to distance ($r/d$), and, hence, the stellar radii. This approach avoids uncertainties in the normalization of observed spectra, which can be important in the vicinity of broad Balmer lines \citep{sergio2020} and, in particular, removes the subjective assignment of the continuum in the UV and near the higher Balmer series lines.


\section{Results}

We replicated the CHORIZOS \citep{maiz2004} analysis presented in Appendix C of \citet{sergio2020} using the new SED results in \teff=\num{15090}$\pm300$\,K for luminosity class (LC) 5 (\logg=4.04) and \teff=\num{15770}$\pm290$\,K for LC 4 (\logg=3.38), with an optical-IR extinction curve \citep{maiz2014} that is very similar to that of \citet{sergio2020}. 
However their best-fit model, with \teff=\num{14000}$\pm500$\,K and \logg=3.5$\pm0.15$, was found to be too cool to match the BD (Figure~\ref{fig:compare}, upper panel, gray line).
We derived a slightly higher temperature, \teff=\num{15300}$\pm200$\,K, and \logg=3.6$\pm0.15$ (assuming N[He/H]=0.1 by number) by fitting the BD and the Balmer lines, which are in good agreement with the CHORIZOS result and that of \citet{sergio2020} (although with a smaller uncertainty in \teff). The SED fit also confirms the small near-IR continuum excess from the disk for a Be disk or an accretion disk. 

The Be+Bstr model of \citet{shenar} with (\teff, \logg, \vt, N[He/H]) of (\num{18000}$\pm$2000\,K, 4.0$\pm$0.3\,dex, 15\,\kms,0.1) and 
(\num{12700}$\pm$2000\,K, 3.0$\pm$0.2\,dex, 2\,\kms, 0.21) for the Be and Bstr components, assuming the Bstr star contributes 55\% of the $V$-band flux and solar metallicity, is also an excellent fit to the G430M data.
Indeed, it is a better fit to the Balmer line cores (which are too deep in the single star model) and to the \ion{He}{I} 4026\,\AA\ line. To some extent, these differences are a consequence of the adopted flux ratio since it is a constraint of the fitting process in the Be+Bstr model, while in the B+BH model, the residuals of the fit can be attributed to the presence of the disk line emission \citep{sergio2020}. \citet{irrgang} proposed a cooler, more helium-rich star, however the spectrum of this lone object is much too cool to fit the BD and this  assumption is not pursued further here.

The UV, on the other hand (Figure 1, lower panel, blue line), demonstrates that
this specific Be+Bstr model is a much poorer fit to the data than the single star model, as exemplified by the \ion{Fe}{III} lines around 1900\,\AA. Since the Be star flux is roughly two to three times that of the stripped star in the UV, and the metal lines in the UV are much stronger than in the optical, the Be star features cannot be easily hidden with a high \vsini.
These differences are largely driven by the adoption of \vt=15\,\kms\ for the Be star, which has a significant impact on the strong saturated lines in the UV. While microturbulent velocities in Be stars are poorly characterized \citep{dunstall2017}, such a high value is clearly ruled out by the UV data, which set an upper limit of \vt=2--3\,\kms.

Using the new data, we can re-assess the Be+Bstr stellar parameters.
For this purpose, we computed an extended grid of composite Be+Bstr TLUSTY models for a range of effective temperatures, microturbulence values, flux ratios, helium abundance (defined as number ratio N[He/H]), and silicon abundance (X[Si], in units of the solar silicon abundance mass faction) using the methods and codes outlined in Section 3. In the following we refer to the effective temperatures of the Be and Bstr stars as $T_{eff}^{Be}$ and $T_{eff}^{Bstr}$ respectively.
As in \citet{shenar}, we initially assume that the Bstr star contributes 55\% of the V-band flux, N[He/H]=0.2, and that \logg for the two components is 4.0 and 3.0 for the Be and Bstr stars, respectively. As demonstrated by \citet{shenar}, and confirmed here, this combination of surface gravities provides an excellent fit to the Balmer lines, that are the primary \logg\ diagnostic. Furthermore, assuming the Be and Bstr stars are the orbital components, the mass ratio strongly constrains the difference in \logg\ to be $\Delta$(\logg)=1.0 dex (also assuming spherical stars).
We also ignore any disk contribution, either continuum or line emission, to the composite spectrum. 

\begin{figure*}[h]
    \centering
    \includegraphics[width=0.49\hsize]{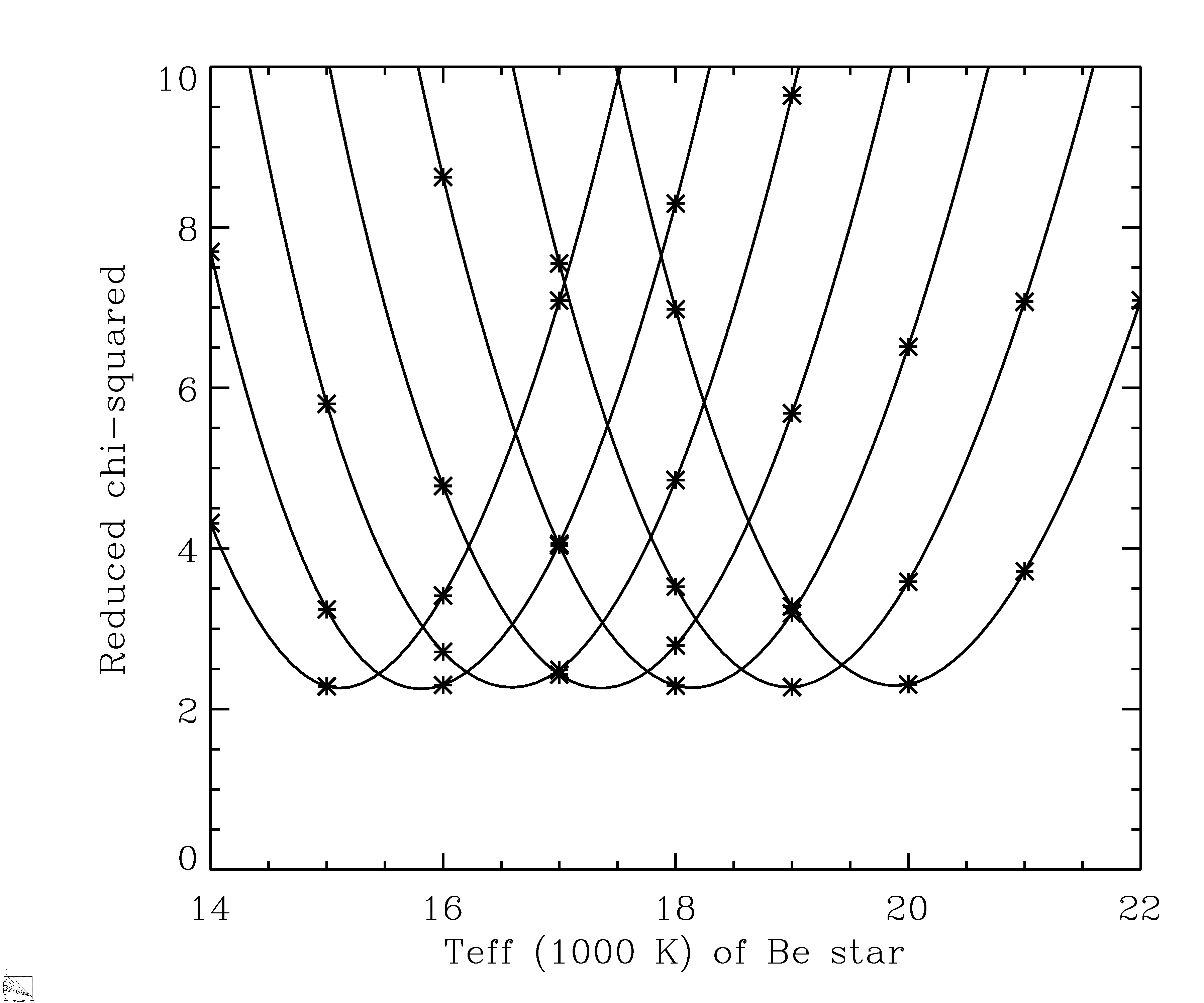}
    \includegraphics[width=0.49\hsize]{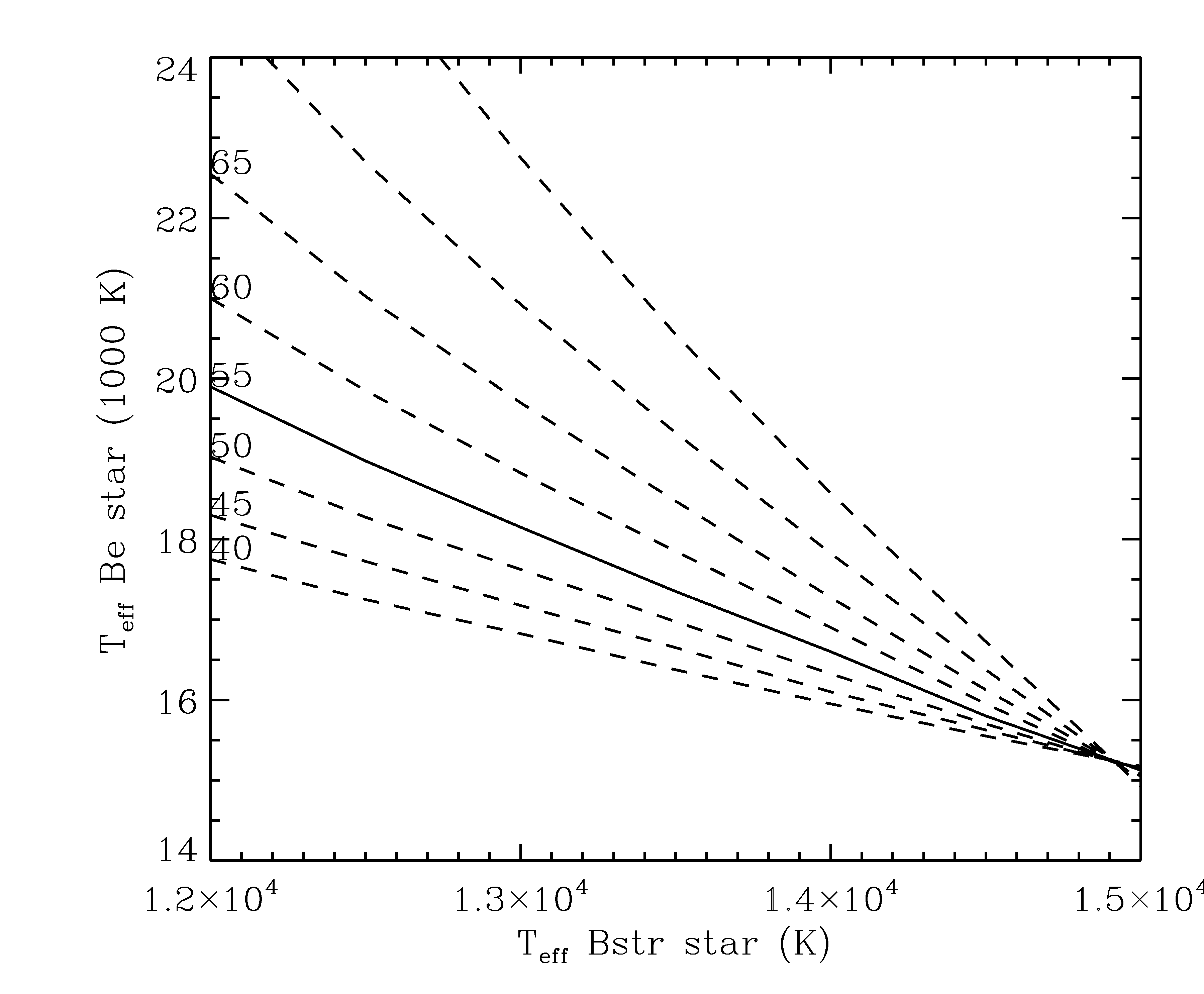}
    \caption{Left-hand panel: Plot showing the reduced-$\chi^2$ of  the fit to the BD region (G430M data). Each curve represents a specific value of $T_{eff}^{Bstr}$ with values ranging from 12000\,K to 15000\,K, right to left, and for a range of $T_{eff}^{Be}$ values. The smooth curves are spline fits to the data points (asterisks). 
    Right-hand panel: Minima of the curves define the relationship between $T_{eff}^{Bstr}$ and $T_{eff}^{Be}$, illustrated by the thick line for a Bstr contribution of 55\% of the flux in the V-band. Thin lines indicate analogous relationships for other percentage flux contributions (labeled).
    }
    \label{fig:teff}
\end{figure*}

For a given choice of $T_{eff}^{Bstr}$ it is straightforward to find the best $T_{eff}^{Be}$ by fitting the BD, as shown in Fig.~\ref{fig:teff} (left panel). 
One can see that the minimum reduced-$\chi^2$ is only weakly dependent on $T_{eff}^{Bstr}$, nevertheless the minima of the curves define the relationship between $T_{eff}^{Bstr}$ and $T_{eff}^{Bstr}$ (Fig.~\ref{fig:teff}, right panel). The slope of this line depends on the adopted flux ratio, as indicated. Therefore if $T_{eff}^{Bstr}$ and the flux ratio can be determined,  $T_{eff}^{Bstr}$ follows from fitting the BD. 

\begin{figure*}[h]
    \centering
  \includegraphics[width=0.49\hsize]{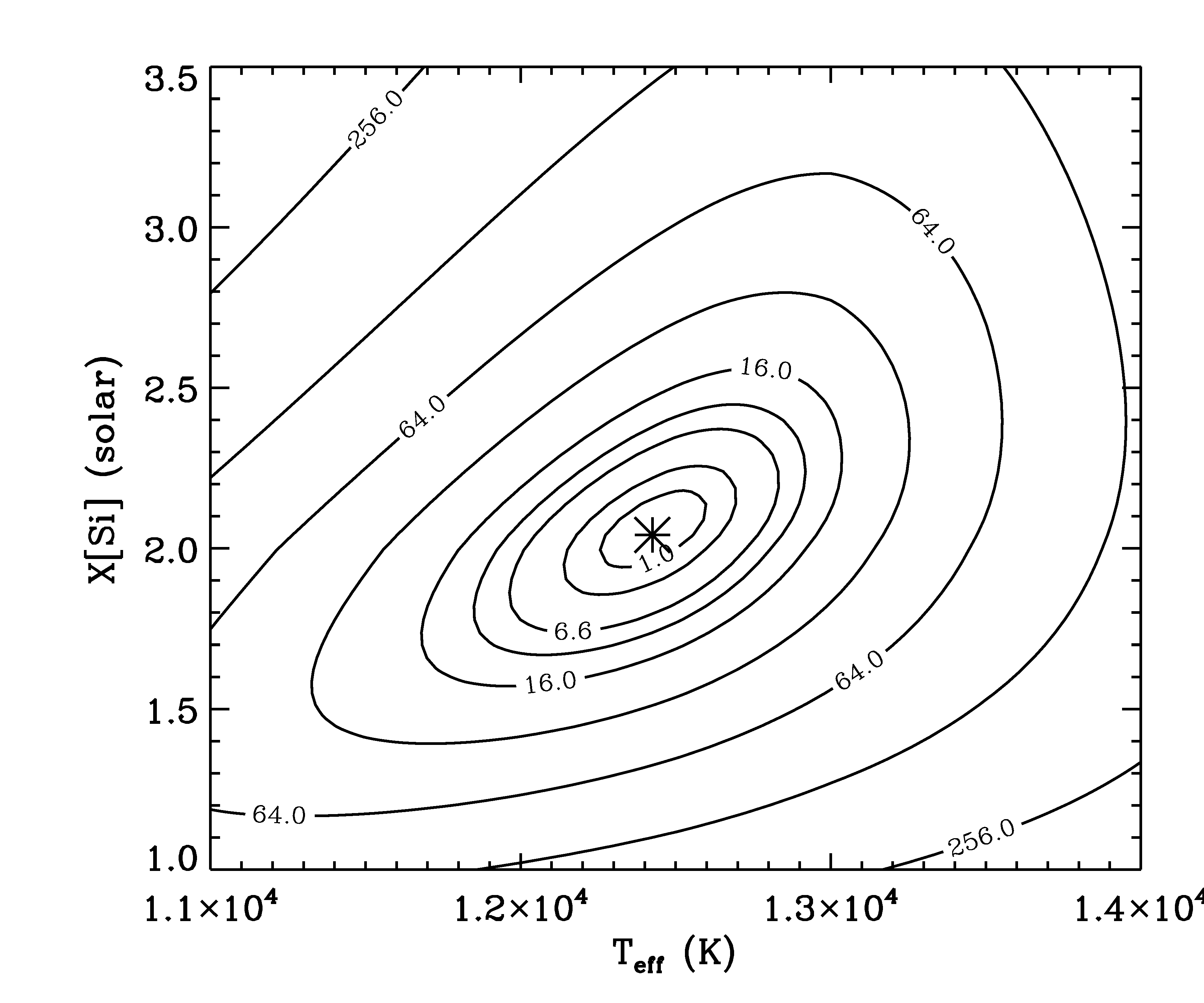}
  \includegraphics[width=0.49\hsize]{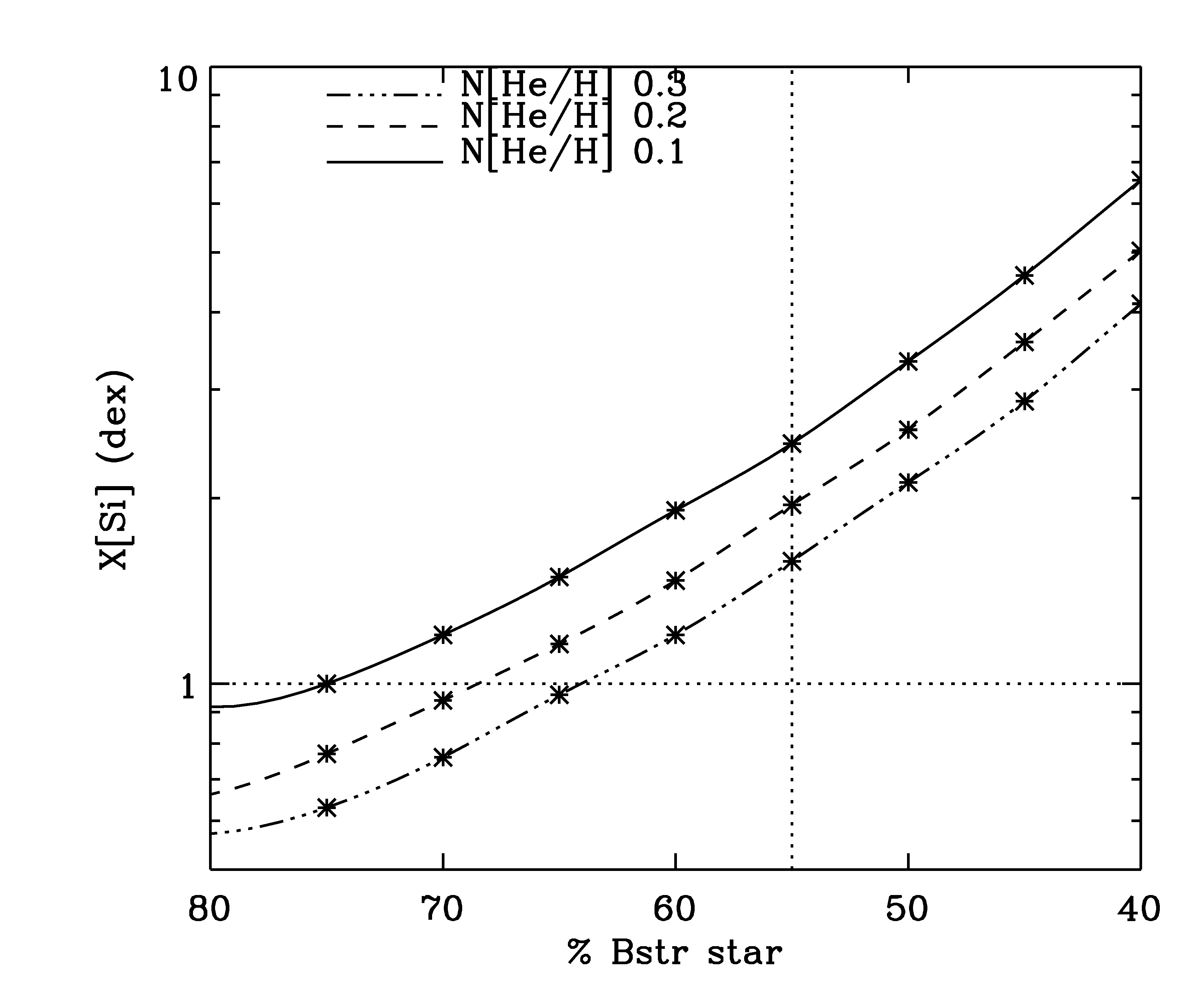}
    \caption{Diagnostic diagrams used to characterize the stellar parameters of the stripped star, and its contribution to the total flux. Left-hand panel: Fit diagram showing reduced-$\chi^2$ contours defining the best fitting $T_{eff}^{Bstr}$ and silicon abundance.
    Right-hand panel: Dependence of best fitting silicon abundance on percentage contribution of the Bstr star to the V-band, and as a function of helium abundance. The horizontal dotted line indicates solar abundance, and the vertical line is the 55\% contribution adopted by \citet{shenar}.
    }
    \label{fig:silicon}
\end{figure*}

\begin{figure}
    \centering
    \includegraphics[width=1.0\hsize]{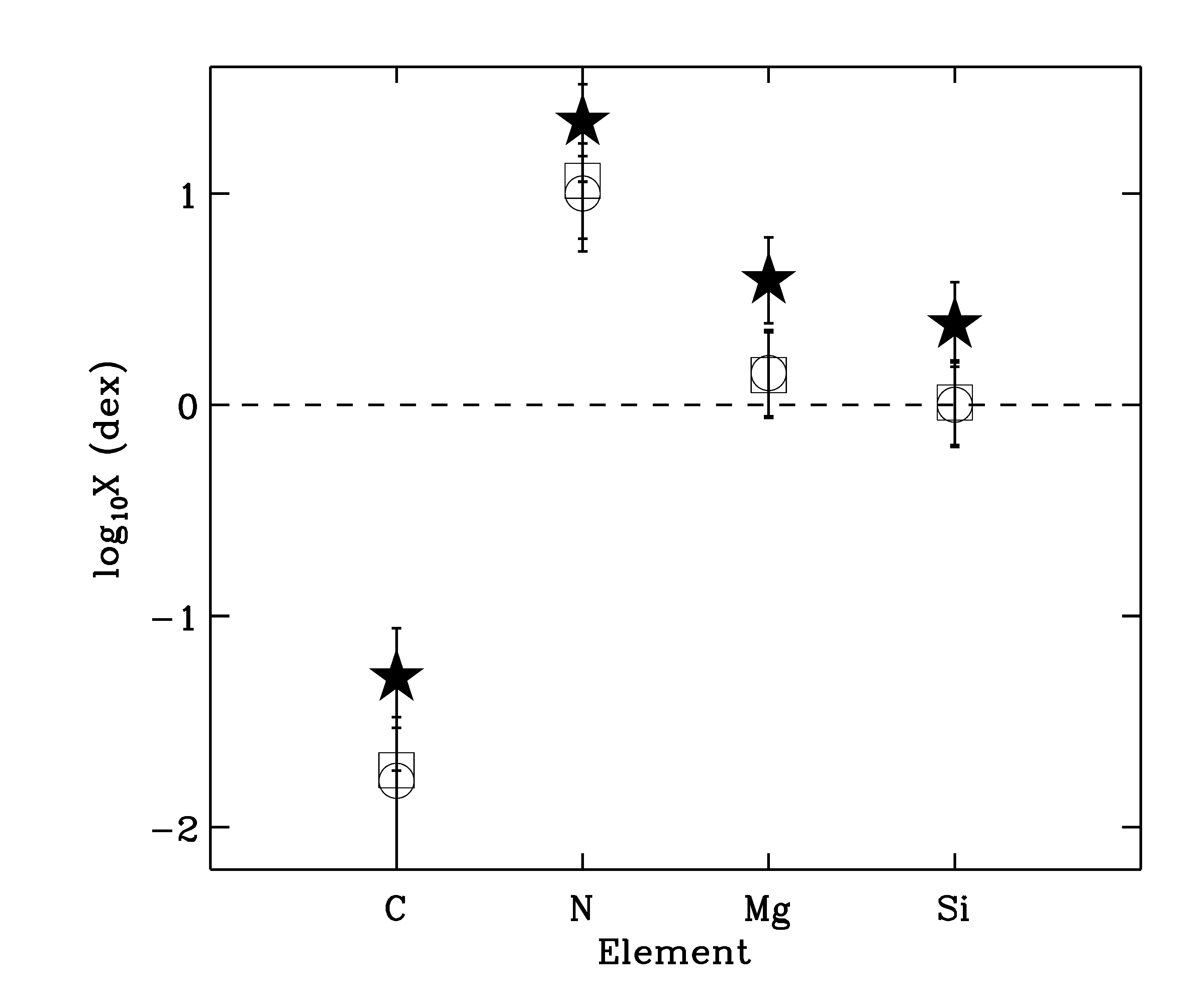}
    \caption{Filled star symbols indicate the abundances obtained for the Bstr model parameters listed in Table \ref{tab:results}.
    The open symbols represent the models that are consistent with a solar Si abundance for different helium abundances, N[He/H]=0.2 (squares) and N[He/H]=0.3 (circles). 
    }
    \label{fig:abundances}
\end{figure}

Adopting the flux ratio from \citet{shenar} we can derive $T_{eff}^{Bstr}$ using the silicon ionization balance, a standard technique in model atmosphere analysis that requires the same silicon abundance from different ionization stages. 
The obvious modification is that the theoretical equivalent widths are measured from the composite spectra. We note that the Be star in this context is simply providing a veiling effect since its metal lines are washed out into the continuum at high \vsini. This is accounted for by combining a Bstr and a Be model, with appropriate weights and parameters, and measuring the theoretical equivalent widths from the resulting spectrum. This approach takes into account small wavelength dependent corrections to the adopted V-band flux ratio that depend on the difference in \teff\ of the two stars.
We use the \ion{Si}{II} doublet lines 4128\,\AA\ and 4131\,\AA, and the \ion{Si}{III} triplet lines at 4553\AA, 4568\AA\ and 4575\,\AA (the 4553\AA\ component is corrected for a blended line of \ion{S}{II}). 
Equivalent widths, with uncertainties, are give in Table \ref{tab:abundances}. We do not use the strong \ion{Si}{II} doublet lines at 6347\,\AA\ and 6371\,\AA\ as the former is blended with a \ion{Mg}{II} doublet and the latter is very discrepant from the other lines. We suspect that this discrepancy may be due to additional veiling of the continuum due to the disk (the system has an IR excess) and, hence, restrict our analysis to the blue optical lines. 
Fit contours in the abundance versus \teff\ plane are illustrated in Fig.\ref{fig:silicon}, from which we find $T_{eff}^{Bstr}$=\num{12500}$\pm$100\,K, and $T_{eff}^{Be}$=\num{18900}$\mp$200\,K from the fit to the BD.
This is in reasonable agreement with \citet{shenar}, though with much reduced uncertainties, given their use of LTE models and their reliance upon \ion{He}{I} to \ion{Mg}{II} line ratios in addition to \ion{Si}{II} to \ion{Si}{III}.
Besides the Si enhancement revealed by Fig.~\ref{fig:silicon}, we find that Mg is enhanced and strong signs of CN processed material in the Bstr star (Table~\ref{tab:abundances} and Fig.~\ref{fig:abundances}).

\begin{table}[h]
    \centering
    \caption{Equivalent widths (EW) of the lines used in this analysis, as measured off the Keck HIRES spectrum discussed in \citet{sergio2020}. The abundances (X) are relative to solar, and are determined for the model with \teff=\num{12500}\,K, \logg=3.0 and for Bstr star contributing 55\% of the total $V$-band flux. Abundance uncertainties reflect observational errors, though uncertainties in \vt\ dominate the overall error budget leading to uncertainties of $\sim$0.1\,dex for C and N, and $\sim$0.2\,dex for Mg and Si. 
    The superscript $^b$ denotes lines that are blended other lines in the data.}
    \begin{tabular}{llrr} \hline \hline
 Ion & Wavelength(\AA) & EW (m\AA) &  X (solar)  \\ \hline
    \ion{Si}{II} &4128.0 &  79.1$\pm$1.0 &   2.39$\pm$0.12\\
    \ion{Si}{II} &4130.8 &  82.4$\pm$1.0 &   1.74$\pm$0.06\\
    \ion{Si}{II} &6347.0$^b$ & 157.0$\pm$3.0 &   6.16$\pm$0.44\\
    \ion{Si}{II} &6371.3 & 118.0$\pm$3.0 &   4.14$\pm$0.42\\
   \ion{Si}{III} &4552.6$^b$ &  11.1$\pm$1.0 &   2.40$\pm$0.43\\
   \ion{Si}{III} &4567.8 &   7.8$\pm$1.0 &   1.98$\pm$0.48\\
   \ion{Si}{III} &4574.7 &   3.7$\pm$0.5 &   1.72$\pm$0.38\\
    \ion{Mg}{II} &4481.2 & 203.6$\pm$3.0 &   3.91$\pm$0.32\\
     \ion{N}{II} &3995.0 &  28.2$\pm$1.0 &  26.93$\pm$2.61\\
     \ion{N}{II} &4447.0 &   9.2$\pm$1.0 &  16.52$\pm$3.41\\
     \ion{N}{II} &4607.1 &  12.4$\pm$1.0 &  35.66$\pm$4.43\\
     \ion{N}{II} &4613.8 &   7.2$\pm$1.0 &  18.73$\pm$4.91\\
     \ion{N}{II} &4630.5 &  20.0$\pm$1.0 &  40.84$\pm$4.29\\
     \ion{C}{II} &3918.9 &  11.2$\pm$1.0 &   0.40$\pm$0.07\\
     \ion{C}{II} &3920.6 &  14.7$\pm$1.0 &   0.33$\pm$0.04\\
     \ion{C}{II} &4267.1 &  28.2$\pm$1.0 &   0.23$\pm$0.02\\
     \ion{C}{II} &6578.0 &   2.8$\pm$1.0 &   0.04$\pm$0.02\\
     \ion{C}{II} &6582.8 &   3.3$\pm$1.0 &   0.08$\pm$0.03\\ \hline
    \end{tabular}
   \label{tab:abundances}
\end{table}

\begin{figure}[h]
   \centering
   \includegraphics[width=1.0\hsize]{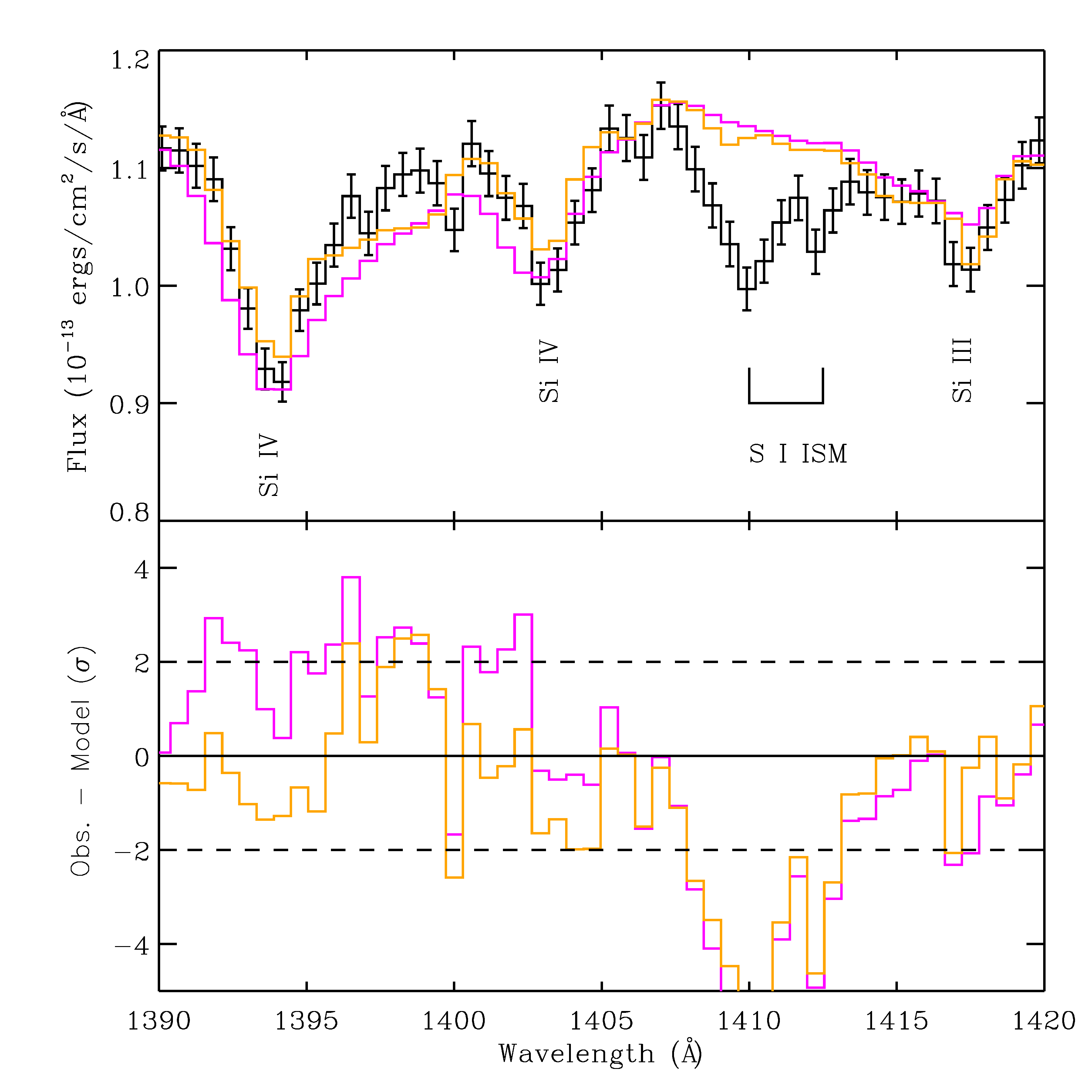}
      \caption{Close-up view of the \ion{Si}{IV} doublet lines at 1393.73\ and 1402.73\,\AA, as well as the \ion{Si}{III} line at 1417.24\,\AA. Color coding is the same as in Figure 1, the observational data now having their statistical error bars added.
      The lower panel illustrates the difference between the the data and the two models in units of $1\sigma$, horizontal dashed lines indicating the $\pm2\sigma$ level.
      }
         \label{fig:doublets}
   \end{figure}

The new solution is illustrated in Fig.~\ref{fig:compare}, now exhibiting a similarly good fit to the UV data (lower panel). We quantify this using the flux errors of the Hubble data to calculate the reduced-$\chi^2$ values of the two models. We exclude the L$_\alpha$ and  \num{2200}\,\AA\ extinction bump regions from this comparison, as these are not sufficiently modeled using a global extinction law (Section 3), and proceed with a piece-wise correction for extinction in the resulting sub-regions of the spectrum. 
We determine reduced-$\chi^2$ values of 4.4 and 3.3 for the Be+Bstr and B+BH models, respectively. Evidently, the latter is slightly better at reproducing the data, the larger value for the Be+Bstr model is due to various metal lines in the Be+Bstr model being slightly too strong and broad. 
However, it is more useful to focus on the features of greatest disagreement in this comparison, such as the \ion{Si}{IV} doublet 1393.73 and 1402.73 \AA\ (Fig.~\ref{fig:compare}, lower panel). In Fig.~\ref{fig:doublets}, we present a close-up view around this feature that clearly shows the superior performance of the B+BH model (i.e., a single star) as the \ion{Si}{IV} doublet is much too strong in the Be+Bstr spectrum.  
This deficiency in the Be+Bstr model is due to the strong contribution of \ion{Si}{IV} and, to a lesser extent \ion{Fe}{III} lines, in the red wing of the 1393.73\,\AA\ component, from the hotter and UV-brighter Be star. We note that these \ion{Si}{IV} resonance lines are so saturated that they are insensitive to even quite large abundance changes and, hence, decreasing the silicon abundance cannot solve this problem. Further, even though the solution illustrated in Fig.~\ref{fig:silicon} implies a silicon abundance for the Bstr star that is roughly twice solar, it is intrinsically weaker in the Bstr star, which is also fainter than the Be star in the UV. Varying \vsini\ of the Be star by $\pm50$\,\kms has a negligible impact on these results, given the spectral resolution. 
 
So far we have assumed that \vt=2\,\kms\ for the Bstr star, but adopting \vt=0\,\kms has negligible effect on the strong UV lines as the thermal Doppler widths of the lines are approximately 3\,\kms\ for Si and 2\,\kms\ for Fe, \citep[see e.g., ][]{mcerlean,smithandhowarth}.
The high silicon abundance referred to above, namely, about twice solar, is clearly at odds with \citet{shenar} who assumed a solar metal abundances for the Bstr star in the determination of its stellar parameters, and in the derivation of the Be/Bstr $V$-band flux ratio. Based on an analogy with the discussion of \vt\ for the Be star, it may seem that increasing this value above 2\,\kms\ to reduce the silicon abundance would be a practical solution. However, this would lead to a value of \vt$\sim$4\,\kms, which is well above the thermal speed of the ions at this low \teff\ and would result in a UV spectrum that is strongly in disagreement with the observations. Also, as discussed in \citet{sergio2020}, the line profiles imply an upper limit of \vt$\sim$2--3\,kms.

In light of the above, we carried out an exploratory calculation in which we varied the flux ratio and the helium abundance of the Bstr star (we adopt grid steps of N[He/H]=0.1, 0.2 and 0.3) to look for solutions with a solar Si abundance. 
It is worth noting that there are some trends aimed at explaining how the constraints drive the solution. For instance,  $T_{eff}^{Bstr}$ is relatively insensitive to helium abundance, but increases a little with increasing relative brightness of the Bstr star, whereas $T_{eff}^{Be}$ increases with greater Bstr star contribution due to the need to compensate for the brighter Bstr star, but a higher Bstr helium abundance leads to a decrease in $T_{eff}^{Be}$ due to the decrease in the BD of the Bstr star.
In Fig.~\ref{fig:silicon}, the right-hand panel shows how the derived silicon abundance varies as a function of flux ratio and N[He/H], implying that a solar Si abundance requires a flux ratio such that the Bstr star contributes roughly 65--70\% of the V-band flux, depending on helium abundance. Specifically, for the N[He/H]=0.2 and 67\% flux contribution, we derive $T_{eff}^{Bstr}$=\num{12700}$\pm$100\,K and $T_{eff}^{Be}$=\num{20875}$\pm$300\,K, while for N[He/H]=0.3 and 64\% flux contribution the corresponding values are \num{12700}$\pm$100\,K and \num{20100}$\pm$250\,K. 
The derived surface abundances are also illustrated in Fig.~\ref{fig:abundances}.
Recalling that the specified  $\Delta$\logg=1.0\,dex ensures that the mass ratio is satisfied for a contribution of 55\%, these solar-Si solutions imply only small changes in this parameter of $\Delta$(\logg)=1.1 and 0.9 dex for the N[He/H]=0.2 and 0.3 cases, respectively. However, the Be star in these cases has a higher temperature than before and the fit to the data is degraded. One possible conclusion is that the Bstr star is enhanced in Si, hence, the flux ratio we inferred is incorrect and undetermined.

Further tests imply that we need to have  $T_{eff}^{Be}\sim$\num{17500}--\num{18000}\,K in order to fit the UV \ion{Si}{IV} lines at these flux ratios, which, in turn implies $T_{eff}^{Bstr}\sim$\num{13000}\,K. While these values are close to those of \citet{shenar}, they are in disagreement with the BD for the above flux ratios. This suggests the need for an investigation of the full parameter space, relaxing the flux ratio and \logg\ constraints (and hence decoupling the solution from the proposed mass ratio).  

\begin{figure}
   \centering
   \includegraphics[width=1.0\hsize]{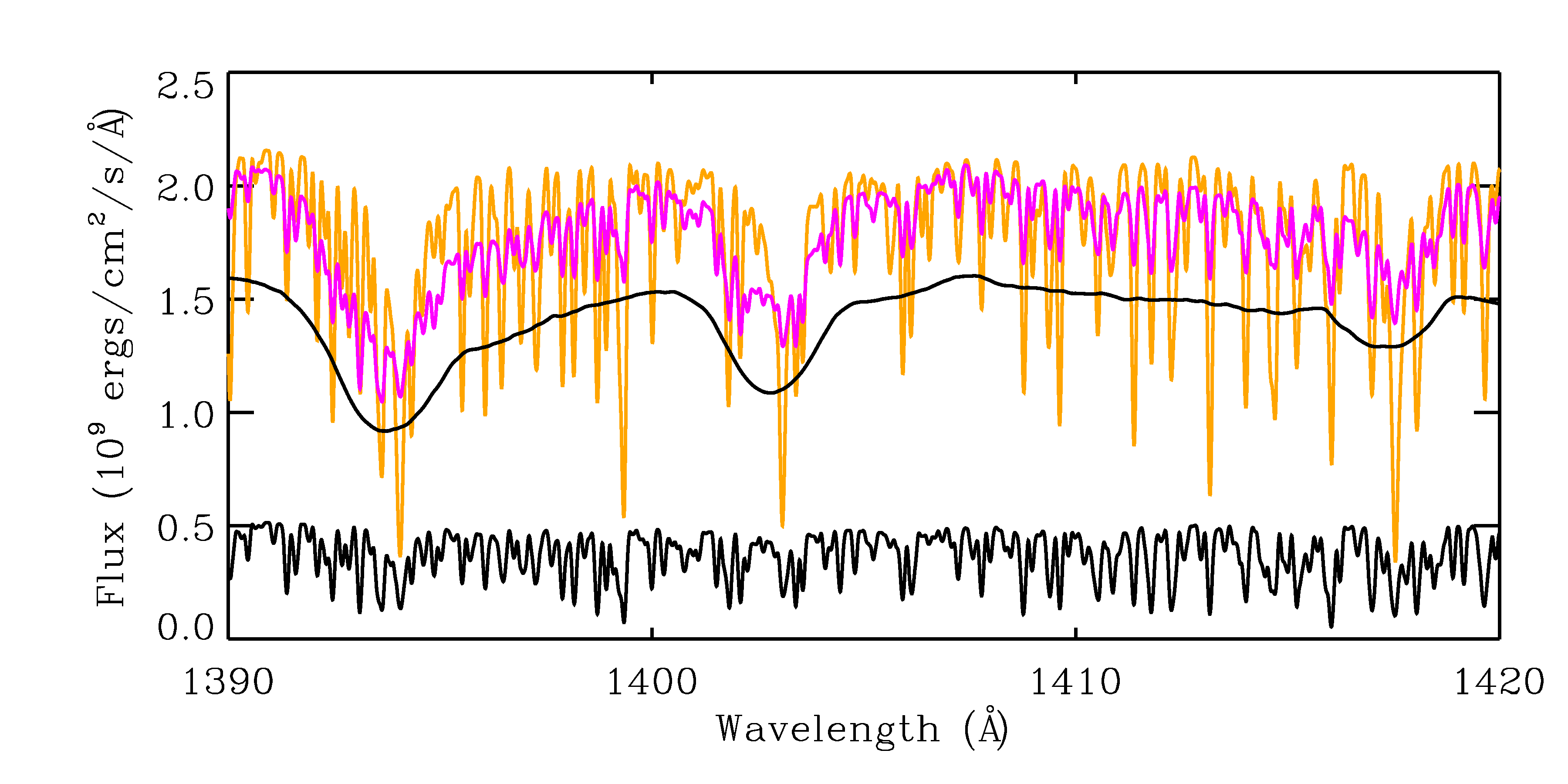}
      \caption{Medium resolution ($R$=\num{20000}) view of the \ion{Si}{IV} doublet for the single star model (orange) and binary model (magenta) with \vt=2\,\kms\ for the Be star. The upper and lower black lines represent the contributions from the Be/Bstr stars, respectively, in the latter scenario.}
         \label{fig:theory}
   \end{figure}

Finally, in this section, in the context of the UV, it is informative to consider higher-resolution predictions for the region containing the \ion{Si}{IV} lines (Figure~\ref{fig:theory}) that clarify the separate contributions to the composite spectrum. This  also serves to emphasize that higher-resolution spectra in the UV, obtainable with STIS E140M or COS G160M modes, can provide strong constraints on the nature of a potential secondary, and critical leverage for spectral disentangling or synthesis.

\begin{table}[]
\caption{Stellar parameters of models discussed here, plus derived quantities as follows; angular radius ($r/d$), stellar radius ($R$), spectroscopic mass ($M_{spec}$), extinction in the $V$ band ($A_V$), and logarithm of the stellar luminosity (log$L$).  $A_V$ is determined from the CHORIZOS analysis, and we adopt a distance of 2.48 kpc from the $Gaia$ EDR3 parallax (see text) in the derivation of the stellar radii and luminosities.}
    \centering
    \begin{tabular}{lccc} 
    \hline
    Model & B+BH & \multicolumn{2}{c}{Be+Bstr} \\
    Component  &  B  & Be  &  Bstr \\ \hline
    \teff\ (K) &\num{15300}$\pm300$ & \num{18900}$\pm200$ & \num{12500}$\pm100$ \\
    \logg  & 3.6$\pm0.2$ & 4.0$\pm0.3$ & 3.0$\pm0.2$ \\
    N[He/H] & 0.10 & 0.10 & 0.20 \\
    \vt\ (\kms) & 2 & 2 & 2 \\
    \vsini\ (\kms)& 8 & 300 & 7 \\ \hline
    $r/d$ (R$_\odot$/kpc) & 2.40$\pm0.04$ &  \multicolumn{2}{c}{2.47$\pm0.04$} \\
    $R$ (R$_\odot$) & 6.0$^{+0.7}_{-0.6}$ & 3.1$^{+0.3}_{-0.3}$ & 4.8$^{+0.5}_{-0.4}$ \\
    $M_{\rm spec}$ (M$_\odot$) & 5.2$^{+1.8}_{-1.4}$ & 3.4$^{+3.5}_{-1.8}$ & 0.8$^{+0.5}_{-0.3}$ \\
    $A_{V}$ (mag) & \multicolumn{3}{c}{1.55$\pm0.03$} \\
    $\log L$ (L$_\odot$)& 3.23$^{+0.09}_{-0.10}$ & 3.04$^{+0.09}_{-0.09}$ & 2.70$^{+0.09}_{-0.09}$ \\
    \hline
    \end{tabular}
     \label{tab:results}
\end{table}

\section{Discussion}

Table \ref{tab:results} lists the stellar parameters derived in the previous section for a Bstr flux contribution of 55\%, as well as the estimated extinction, angular radius, stellar radius, and spectroscopic masses for both B+BH and Be+Bstr models.

In deriving stellar radii, we use the Gaia EDR3 results \citep{gaiacollaboration2020gaia} for the LB-1 parallax of $\varpi=0.3592\pm0.0296$ mas, notably corrected as recommended by the recipe from \citet{edr3biaslindegren}, which depends on magnitude, color, and ecliptic distance, and that for our target yields a zero point of $-$0.0511~mas. Such a correction does not include the effect of the covariance for small angular distances seen in the LMC data of \citet{edr3astrlindegren}, namely, the checkered pattern, and to account for it, we add 0.0260~mas in quadrature to the parallax uncertainty\footnote{This is a conservative estimate based on the measurements for quasars of \citet{edr3astrlindegren}. It may be possible to refine it in the future using further analysis (Ma{\'\i}z Apell\'aniz et al. in preparation).}, resulting in $\varpi=0.4103\pm0.0394$ mas. 
Using the OB star prior of \citet{maiz2001,maiz2008} this leads to a distance of $2.48^{+0.27}_{-0.22}$\,kpc, consistent with $2.20^{+0.49}_{-0.35}$\,kpc estimated in \citet{sergio2020} using $Gaia$ DR2 data, though with a smaller uncertainty. 
The EDR3 data for LB-1 are now based on 26 transits, compared to 14 in DR2, and following the discussion in Appendix D of \citet{sergio2020}, these new data also do not display evidence for orbital motion of the B-star \citep[see also the discussion in][]{eldridge2020}. 
The {\tt ruwe} parameter of 1.22 still indicates a clean astrometric fit, while the image parameter determination quality flags, {\tt ipd\_multi\_peak} and {\tt ipd\_odd\_win,} are both 0, which is consistent with the PSF from the WFC3/IR image. However, while the goodness of fit parameter {\tt ipd\_gof\_harmonic\_amplitude} is on the high side at 0.09, this is not reflected in the WFC3/IR images mentioned in Section 2, which have negligible ellipticity. As discussed in detail in Appendix D of \citet{sergio2020},  we attribute the puzzling lack of evidence for the orbital motion of the system to it being aligned almost edge-on and the particular circumstances of its orientation with respect to the sun and its proper motion vector.

\begin{figure}
   \centering
   \includegraphics[width=0.9\hsize]{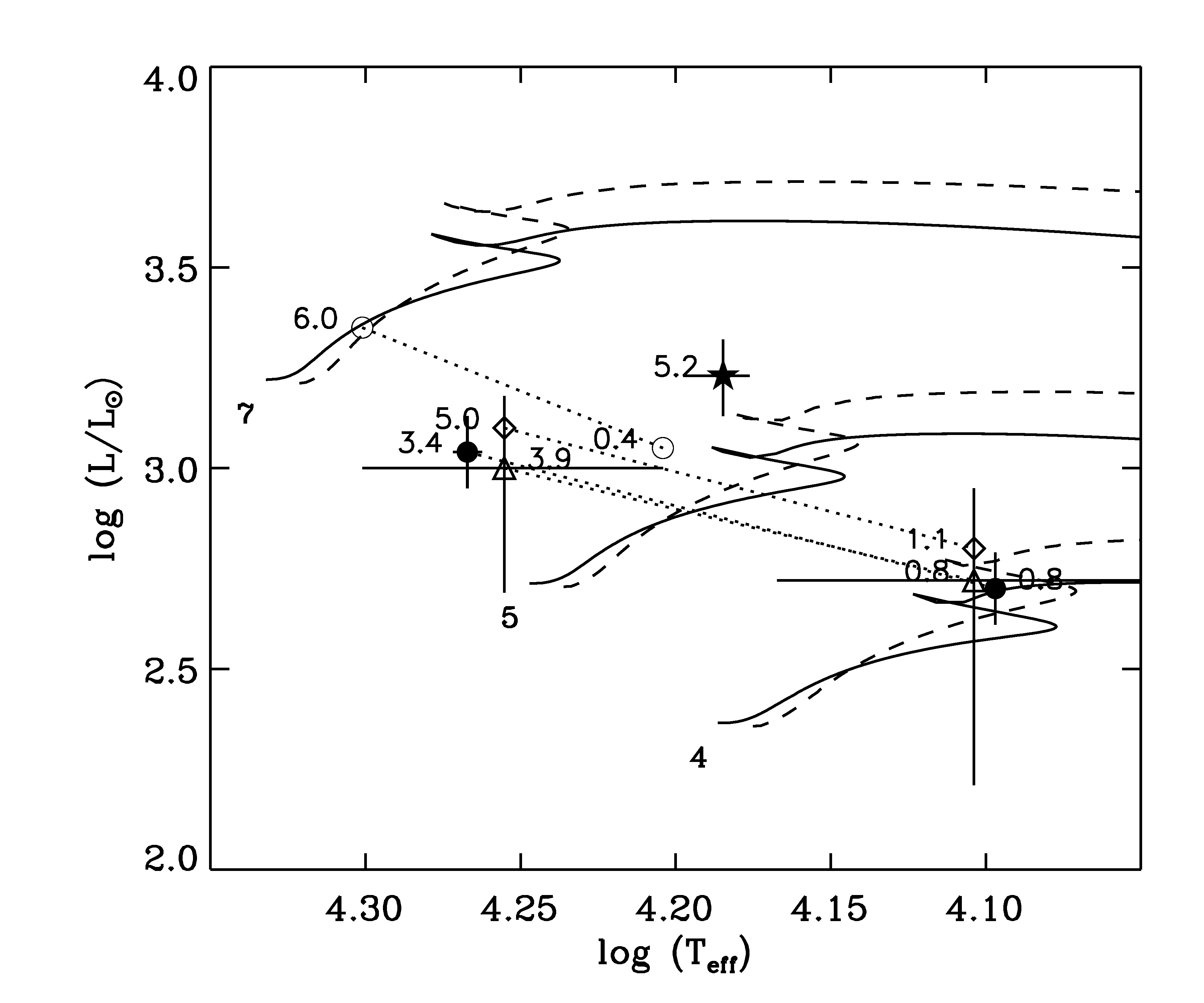}
      \caption{HRD for the components of LB-1 as derived here (with error bars). The single star solution is the filled star, while the binary solution components are filled circles joined by a dotted line. The Be star, or broad lined star, is the hotter of these two components.
      The \citet{shenar} parameters are indicated (diamonds), as are our estimates using our updated values of extinction and distance (triangles). Also indicated  are the positions for HR\,6819 (open circles) as determined by \citet{bodensteiner2020}. For context the evolutionary tracks are the non-rotating (solid lines) and rotating (dashed lines) single star models from \citet{ekstrom}. Symbols are labeled with their spectroscopic masses, and each track is labeled with its initial mass. }
         \label{fig:hrd}
   \end{figure}

The luminosity of the LB-1 objects are now tightly constrained by the Hubble flux-calibrated spectrum, the well-determined extinction law, and the $Gaia$ EDR3 parallax.
As the HRD in Fig.~\ref{fig:hrd} demonstrates, for the Be+Bstr solution, the Be star component is close to the zero age main sequence (ZAMS), which is  not typical for classical Be stars \citep{fabregat}.  
This characteristic is shared with the Be star proposed by \citet{bodensteiner2020} for the system HR\,6819.
Further, as discussed above, solutions encompassing a solar Si abundance predict an even smaller and hotter Be star, exacerbating this discrepancy. However these models cannot reproduce the observed UV spectrum, which implies the need for a cooler Be star (\teff$\sim$\num{17500}--\num{18000}\,K), and hence a hotter silicon-rich Bstr star (\teff$\sim$\num{13000}). This family of solutions effectively relaxes the constraints on the flux and mass ratio and indicates the need for a more complete exploration of the available parameter space,  now including the UV.  This task is beyond the scope of this paper but would be a useful check to ascertain that the Balmer emission is indeed a measure of the reflex velocity of the companion to the narrow-lined star.

The Be+Bstr model also results in a rather small spectroscopic mass for the Be star, namely, 3.4 M$_\odot$, which is also much too small to be considered a classical Be star of this temperature \citep[see][]{rivinius2013}. However this estimate assumes spherical symmetry, which may not be correct. As discussed by \citet{fre05}, oblateness can lead to estimated gravities being up to 0.4 dex lower than those found at the stellar pole. In turn, this can lead to a significant underestimation of the stellar mass. 
Additionally, if the Gaia parallax were incorrect, then a distance of $\sim$3.5\,kpc is required to move the Be star into the vicinity of the end of the main sequence. The Bstr star would then have a mass of $\sim$1.7\,$M_{\odot}$.

Figure \ref{fig:hrd} also demonstrates good agreement between spectroscopic and evolutionary masses of the B-type star in the B+BH solution, demonstrating an improvement on \citet{sergio2020}. The current mass implies a potential BH mass of $\sim21^{+9}_{-8}\,{\rm M}_\odot$ using the revised mass ratio of $5.1\pm0.1$ \citep{liu2020}. However, if the very low \vsini\ is a consequence of binary interaction the agreement of the spectroscopic mass with single-star evolutionary tracks may be fortuitous. Nevertheless, the distance discussed here restricts the upper limit on the X-ray luminosity of LB-1 to $\sim$6x10$^{30}$\,erg\,s$^{-1}$, which is consistent with the faintest known quiescent BH accretion disks \citep{armas,ribo}. 

While the Be nature of the broad lined star has been attributed to the presence of a disk, as implied by the characteristic Balmer emission lines and IR excess, the SED displays no evidence for a second BD or emission in the \ion{Mg}{II}~2800~\AA\ doublet, as is often observed in classical Be stars \citep{cochetti,slettebak}.  \citet{liu2020} have discussed the Balmer and Paschen emission line spectrum at length, and detect emission wings to a velocity of $\pm250$ \kms, that for a Keplerian disk measure the projected velocity of the inner edge of the disk. The radius and mass of the Be star in Table 3 lead to values of $\sin i\sim0.52$ and a dynamical mass of 12.8 M$_\odot$ \citep[assuming $M\sin^3i=1.78$ from][]{shenar}, adding further tension with the estimated spectroscopic mass of the Be-type star. In order to match the spectroscopic mass one requires $\sin i\sim 0.8$, supporting the argument that the system is viewed almost edge-on. Obviously, in the context of the B+BH scenario in which the emission arises from an accretion disk around the BH, the above argument does not apply. However, in this case, the small line widths tend to favour a low inclination angle for the accretion disk.

\section{Conclusions}

The Hubble data enable more precise stellar parameters for the LB-1 system within the B+BH and Be+Bstr models. The B+BH (single star) solution is a better fit to the UV, although the Be+Bstr star model remains superior in the optical. We indicate how the B+BH UV performance can be improved by a combination of a fainter Be star and more helium-rich Bstr star. However, the Be is already close to the ZAMS, unusual for a classical Be star, and this solution would shift it even closer to the ZAMS.  We find enhanced Si and Mg in the Bstr star, although with a non-solar abundance ratio (Table~\ref{tab:abundances}) and signs of extreme CN processing.
In the B+BH model, the B-type star's position in the HRD now provides better agreement between evolutionary and spectroscopic masses, resolving the tension that previously existed between these masses \citep{sergio2020}. However, the very low \vsini\, and likely low $v_e$ hints at an evolutionary path that involves binary interaction.

Indeed, it is difficult to reconcile the properties of the LB-1 with any of the current evolutionary scenarios \citep[see also the discussion by][]{liu2020}. We find that higher-resolution UV spectra would serve as a powerful diagnostic for characterizing the nature of a potential companion to the narrow-lined star. 

\begin{acknowledgements}
This work was supported by the Spanish Ministry of Science and Innovation through grants PGC2018-091\,3741-B-C22 and PGC2018-095\,049-B-C22,
the European Regional Development Fund under grants AYA2017-83216-P, AYA2017-86389-P, ProID2017-01011-5 and the Canarian Agency for Research, Innovation and Information Society.
TMD and JIGH acknowledge support via the Ram\'on y Cajal Fellowships RYC-2015-18148 and RYC-2013-14875 respectively.
A.I.\ acknowledges funding by the Deutsche For\-schungs\-gemeinschaft (DFG) through grant HE1356/70-1. RB and SD acknowledge support from NASA through grant number O2064 from the Space Telescope Science Institute, which is operated by AURA, Inc., under NASA contract NAS 5-26555.
This work has made use of data from the European Space Agency (ESA) mission
{\it Gaia} (\url{https://www.cosmos.esa.int/gaia}), processed by the {\it Gaia}
Data Processing and Analysis Consortium (DPAC,
\url{https://www.cosmos.esa.int/web/gaia/dpac/consortium}). Funding for the DPAC
has been provided by national institutions, in particular those
participating in the {\it Gaia} Multilateral Agreement.

\end{acknowledgements}

%
%
\bibliographystyle{aa}
\bibliography{references.bib}

\begin{thebibliography}{35}
\expandafter\ifx\csname natexlab\endcsname\relax\def\natexlab#1{#1}\fi

\bibitem[{{Abdul-Masih} {et~al.}(2020){Abdul-Masih}, {Banyard}, {Bodensteiner},
  {Bordier}, {Bowman}, {Dsilva}, {Fabry}, {Hawcroft}, {Mahy}, {Marchant},
  {Raskin}, {Reggiani}, {Shenar}, {Tkachenko}, {Van Winckel}, {Vermeylen}, \&
  {Sana}}]{abdulmasih}
{Abdul-Masih}, M., {Banyard}, G., {Bodensteiner}, J., {et~al.} 2020, \nat, 580,
  E11

\bibitem[{{Armas Padilla} {et~al.}(2014){Armas Padilla}, {Wijnands},
  {Degenaar}, {Mu{\~n}oz-Darias}, {Casares}, \& {Fender}}]{armas}
{Armas Padilla}, M., {Wijnands}, R., {Degenaar}, N., {et~al.} 2014, \mnras,
  444, 902

\bibitem[{{Bodensteiner} {et~al.}(2020){Bodensteiner}, {Shenar}, {Mahy},
  {Fabry}, {Marchant}, {Abdul-Masih}, {Banyard}, {Bowman}, {Dsilva}, {Frost},
  {Hawcroft}, {Reggiani}, \& {Sana}}]{bodensteiner2020}
{Bodensteiner}, J., {Shenar}, T., {Mahy}, L., {et~al.} 2020, \aap, 641, A43

\bibitem[{{Bohlin} \& {Deustua}(2019)}]{bohlinIR}
{Bohlin}, R.~C. \& {Deustua}, S.~E. 2019, \aj, 157, 229

\bibitem[{{Bohlin} {et~al.}(2019){Bohlin}, {Deustua}, \& {de
  Rosa}}]{bohlin2019}
{Bohlin}, R.~C., {Deustua}, S.~E., \& {de Rosa}, G. 2019, \aj, 158, 211

\bibitem[{{Bohlin} {et~al.}(2014){Bohlin}, {Gordon}, \&
  {Tremblay}}]{bohlin2014}
{Bohlin}, R.~C., {Gordon}, K.~D., \& {Tremblay}, P.~E. 2014, \pasp, 126, 711

\bibitem[{{Bohlin} {et~al.}(2020){Bohlin}, {Hubeny}, \& {Rauch}}]{bohlin2020}
{Bohlin}, R.~C., {Hubeny}, I., \& {Rauch}, T. 2020, \aj, 160, 21

\bibitem[{{Cochetti} {et~al.}(2020){Cochetti}, {Zorec}, {Cidale}, {Arias},
  {Aidelman}, {Torres}, {Fr{\'e}mat}, \& {Granada}}]{cochetti}
{Cochetti}, Y.~R., {Zorec}, J., {Cidale}, L.~S., {et~al.} 2020, \aap, 634, A18

\bibitem[{{Dunstall} {et~al.}(2011){Dunstall}, {Brott}, {Dufton}, {Lennon},
  {Evans}, {Smartt}, \& {Hunter}}]{dunstall2017}
{Dunstall}, P.~R., {Brott}, I., {Dufton}, P.~L., {et~al.} 2011, \aap, 536, A65

\bibitem[{{Ekstr{\"o}m} {et~al.}(2012){Ekstr{\"o}m}, {Georgy}, {Eggenberger},
  {Meynet}, {Mowlavi}, {Wyttenbach}, {Granada}, {Decressin}, {Hirschi},
  {Frischknecht}, {Charbonnel}, \& {Maeder}}]{ekstrom}
{Ekstr{\"o}m}, S., {Georgy}, C., {Eggenberger}, P., {et~al.} 2012, \aap, 537,
  A146

\bibitem[{{Eldridge} {et~al.}(2020){Eldridge}, {Stanway}, {Breivik}, {Casey},
  {Steeghs}, \& {Stevance}}]{eldridge2020}
{Eldridge}, J.~J., {Stanway}, E.~R., {Breivik}, K., {et~al.} 2020, \mnras, 495,
  2786

\bibitem[{{Fabregat} \& {Torrej{\'o}n}(2000)}]{fabregat}
{Fabregat}, J. \& {Torrej{\'o}n}, J.~M. 2000, \aap, 357, 451

\bibitem[{{Fitzpatrick} {et~al.}(2019){Fitzpatrick}, {Massa}, {Gordon},
  {Bohlin}, \& {Clayton}}]{fitzpatrick2019}
{Fitzpatrick}, E.~L., {Massa}, D., {Gordon}, K.~D., {Bohlin}, R., \& {Clayton},
  G.~C. 2019, \apj, 886, 108

\bibitem[{{Fr{\'e}mat} {et~al.}(2005){Fr{\'e}mat}, {Zorec}, {Hubert}, \&
  {Floquet}}]{fre05}
{Fr{\'e}mat}, Y., {Zorec}, J., {Hubert}, A.-M., \& {Floquet}, M. 2005, \aap,
  440, 305

\bibitem[{{Gaia Collaboration} {et~al.}(2020){Gaia Collaboration}, Brown,
  Vallenari, Prusti, de~Bruijne, Babusiaux, \&
  Biermann}]{gaiacollaboration2020gaia}
{Gaia Collaboration}, Brown, A. G.~A., Vallenari, A., {et~al.} 2020, Gaia Early
  Data Release 3: Summary of the contents and survey properties

\bibitem[{{Hubeny}(1988)}]{hubeny1988}
{Hubeny}, I. 1988, Computer Physics Communications, 52, 103

\bibitem[{{Hubeny} \& {Lanz}(1995)}]{hubeny1995}
{Hubeny}, I. \& {Lanz}, T. 1995, \apj, 439, 875

\bibitem[{{Irrgang} {et~al.}(2020){Irrgang}, {Geier}, {Kreuzer}, {Pelisoli}, \&
  {Heber}}]{irrgang}
{Irrgang}, A., {Geier}, S., {Kreuzer}, S., {Pelisoli}, I., \& {Heber}, U. 2020,
  \aap, 633, L5

\bibitem[{{Lanz} \& {Hubeny}(2007)}]{lanz2007}
{Lanz}, T. \& {Hubeny}, I. 2007, \apjs, 169, 83

\bibitem[{{Lindegren} {et~al.}(2020{\natexlab{a}}){Lindegren}, {Bastian},
  {Biermann}, {Bombrun}, {de Torres}, {Gerlach}, {Geyer}, {Hern{\'a}ndez},
  {Hilger}, {Hobbs}, {Klioner}, {Lammers}, {McMillan}, {Ramos-Lerate},
  {Steidelm{\"u}ller}, {Stephenson}, \& {van Leeuwen}}]{edr3biaslindegren}
{Lindegren}, L., {Bastian}, U., {Biermann}, M., {et~al.} 2020{\natexlab{a}},
  arXiv e-prints, arXiv:2012.01742

\bibitem[{{Lindegren} {et~al.}(2020{\natexlab{b}}){Lindegren}, {Klioner},
  {Hern{\'a}ndez}, {Bombrun}, {Ramos-Lerate}, {Steidelm{\"u}ller}, {Bastian},
  {Biermann}, {de Torres}, {Gerlach}, {Geyer}, {Hilger}, {Hobbs}, {Lammers},
  {McMillan}, {Stephenson}, {Casta{\~n}eda}, {Davidson}, {Fabricius},
  {Gracia-Abril}, {Portell}, {Rowell}, {Teyssier}, {Torra}, {Bartolom{\'e}},
  {Clotet}, {Garralda}, {Gonz{\'a}lez-Vidal}, {Torra}, {Abbas}, {Altmann},
  {Anglada Varela}, {Balaguer-N{\'u}{\~n}ez}, {Balog}, {Barache}, {Becciani},
  {Bernet}, {Bertone}, {Bianchi}, {Bouquillon}, {Brown}, {Bucciarelli},
  {Busonero}, {Butkevich}, {Buzzi}, {Cancelliere}, {Carlucci}, {Charlot},
  {Cioni}, {Crosta}, {Crowley}, {del Peloso}, {del Pozo}, {Drimmel}, {Esquej},
  {Fienga}, {Fraile}, {Gai}, {Garcia-Reinaldos}, {Guerra}, {Hambly}, {Hauser},
  {Jan{\ss}en}, {Jordan}, {Kostrzewa-Rutkowska}, {Lattanzi}, {Liao}, {Licata},
  {Lister}, {L{\"o}ffler}, {Marchant}, {Masip}, {Mignard}, {Mints}, {Molina},
  {Mora}, {Morbidelli}, {Murphy}, {Pagani}, {Panuzzo}, {Pe{\~n}alosa Esteller},
  {Poggio}, {Re Fiorentin}, {Riva}, {Sagrist{\`a} Sell{\'e}s}, {Sanchez
  Gimenez}, {Sarasso}, {Sciacca}, {Siddiqui}, {Smart}, {Souami}, {Spagna},
  {Steele}, {Taris}, {Utrilla}, {van Reeven}, \&
  {Vecchiato}}]{edr3astrlindegren}
{Lindegren}, L., {Klioner}, S.~A., {Hern{\'a}ndez}, J., {et~al.}
  2020{\natexlab{b}}, arXiv e-prints, arXiv:2012.03380

\bibitem[{{Liu} {et~al.}(2019){Liu}, {Zhang}, {Howard}, {Bai}, {Lu}, {Soria},
  {Justham}, {Li}, {Zheng}, {Wang}, {Belczynski}, {Casares}, {Zhang}, {Yuan},
  {Dong}, {Lei}, {Isaacson}, {Wang}, {Bai}, {Shao}, {Gao}, {Wang}, {Niu},
  {Cui}, {Zheng}, {Mu}, {Zhang}, {Wang}, {Heger}, {Qi}, {Liao}, {Lattanzi},
  {Gu}, {Wang}, {Wu}, {Shao}, {Shen}, {Wang}, {Bregman}, {Di Stefano}, {Liu},
  {Han}, {Zhang}, {Wang}, {Ren}, {Zhang}, {Zhang}, {Wang}, {Cabrera-Lavers},
  {Corradi}, {Rebolo}, {Zhao}, {Zhao}, {Chu}, \& {Cui}}]{liu}
{Liu}, J., {Zhang}, H., {Howard}, A.~W., {et~al.} 2019, \nat, 575, 618

\bibitem[{{Liu} {et~al.}(2020){Liu}, {Zheng}, {Soria}, {Aceituno}, {Zhang},
  {Lu}, {Wang}, {Hamann}, {Oskinova}, {Ramachandran}, {Yuan}, {Bai}, {Wang},
  {McKee}, {Wu}, {Wang}, {Lattanzi}, {Belczynski}, {Casares}, {Gonz{\'a}lez
  Hern{\'a}ndez}, \& {Rebolo}}]{liu2020}
{Liu}, J., {Zheng}, Z., {Soria}, R., {et~al.} 2020, \apj, 900, 42

\bibitem[{{Ma{\'\i}z Apell{\'a}niz}(2001)}]{maiz2001}
{Ma{\'\i}z Apell{\'a}niz}, J. 2001, \aj, 121, 2737

\bibitem[{{Ma{\'\i}z Apell{\'a}niz}(2004)}]{maiz2004}
{Ma{\'\i}z Apell{\'a}niz}, J. 2004, \pasp, 116, 859

\bibitem[{{Ma{\'\i}z Apell{\'a}niz} {et~al.}(2008){Ma{\'\i}z Apell{\'a}niz},
  {Alfaro}, \& {Sota}}]{maiz2008}
{Ma{\'\i}z Apell{\'a}niz}, J., {Alfaro}, E.~J., \& {Sota}, A. 2008, arXiv
  e-prints, arXiv:0804.2553

\bibitem[{{Ma{\'\i}z Apell{\'a}niz} {et~al.}(2014){Ma{\'\i}z Apell{\'a}niz},
  {Evans}, {Barb{\'a}}, {Gr{\"a}fener}, {Bestenlehner}, {Crowther},
  {Garc{\'\i}a}, {Herrero}, {Sana}, {Sim{\'o}n-D{\'\i}az}, {Taylor}, {van
  Loon}, {Vink}, \& {Walborn}}]{maiz2014}
{Ma{\'\i}z Apell{\'a}niz}, J., {Evans}, C.~J., {Barb{\'a}}, R.~H., {et~al.}
  2014, \aap, 564, A63

\bibitem[{{McErlean} {et~al.}(1998){McErlean}, {Lennon}, \&
  {Dufton}}]{mcerlean}
{McErlean}, N.~D., {Lennon}, D.~J., \& {Dufton}, P.~L. 1998, \aap, 329, 613

\bibitem[{{Rib{\'o}} {et~al.}(2017){Rib{\'o}}, {Munar-Adrover}, {Paredes},
  {Marcote}, {Iwasawa}, {Mold{\'o}n}, {Casares}, {Migliari}, \&
  {Paredes-Fortuny}}]{ribo}
{Rib{\'o}}, M., {Munar-Adrover}, P., {Paredes}, J.~M., {et~al.} 2017, \apjl,
  835, L33

\bibitem[{{Rivinius} {et~al.}(2020){Rivinius}, {Baade}, {Hadrava}, {Heida}, \&
  {Klement}}]{rivinius}
{Rivinius}, T., {Baade}, D., {Hadrava}, P., {Heida}, M., \& {Klement}, R. 2020,
  \aap, 637, L3

\bibitem[{{Rivinius} {et~al.}(2013){Rivinius}, {Carciofi}, \&
  {Martayan}}]{rivinius2013}
{Rivinius}, T., {Carciofi}, A.~C., \& {Martayan}, C. 2013, \aapr, 21, 69

\bibitem[{{Shenar} {et~al.}(2020){Shenar}, {Bodensteiner}, {Abdul-Masih},
  {Fabry}, {Mahy}, {Marchant}, {Banyard}, {Bowman}, {Dsilva}, {Hawcroft},
  {Reggiani}, \& {Sana}}]{shenar}
{Shenar}, T., {Bodensteiner}, J., {Abdul-Masih}, M., {et~al.} 2020, \aap, 639,
  L6

\bibitem[{{Sim{\'o}n-D{\'\i}az} {et~al.}(2020){Sim{\'o}n-D{\'\i}az}, {Ma{\'\i}z
  Apell{\'a}niz}, {Lennon}, {Gonz{\'a}lez Hern{\'a}ndez}, {Allende Prieto},
  {Castro}, {de Burgos}, {Dufton}, {Herrero}, {Toledo-Padr{\'o}n}, \&
  {Smartt}}]{sergio2020}
{Sim{\'o}n-D{\'\i}az}, S., {Ma{\'\i}z Apell{\'a}niz}, J., {Lennon}, D.~J.,
  {et~al.} 2020, \aap, 634, L7

\bibitem[{{Slettebak}(1994)}]{slettebak}
{Slettebak}, A. 1994, \apjs, 94, 163

\bibitem[{{Smith} \& {Howarth}(1998)}]{smithandhowarth}
{Smith}, K.~C. \& {Howarth}, I.~D. 1998, \mnras, 299, 1146

\end{thebibliography}

\end{document}